\documentclass[useAMS,usenatbib]{mn2e}

\title[The Independency of Stellar Mass Loss Rates on stellar X-ray Luminosity and Activity Level]
{The Independency of Stellar Mass Loss Rates on stellar X-ray Luminosity and Activity Level based on Solar X-ray flux and Solar Wind Observations}
\author[O. Cohen]{O. Cohen$^{1}$\thanks{E-mail:
ocohen@cfa.harvard.edu (OC)}\\
$^{1}$Harvard-Smithsonian Center for Astrophysics, 60 Garden St. Cambridge, MA 02138, USA}

\usepackage{graphicx} 
\begin{document}

\date{}

\pagerange{\pageref{firstpage}--\pageref{lastpage}} \pubyear{2002}

\maketitle

\label{firstpage}


\begin{abstract}

Stellar mass loss rates are an important input ingredient for stellar evolution models since they determine stellar evolution parameters such as stellar spin-down and increase in stellar luminosity through the lifetime of a star. Due to the lack of direct observations of stellar winds from Sun-like stars stellar X-ray luminosity and stellar level of activity are commonly used as a proxy for estimating stellar mass loss rates. However, such an intuitive activity --- mass loss rate relation is not well defined. In this paper, I study the mass loss rate of the Sun as a function of its activity level. I compare in-situ solar wind measurements with the solar activity level represented by the solar X-ray flux. I find no clear dependency of the solar mass flux on solar X-ray flux. Instead, the solar mass loss rate is scattered around an average value of $2\cdot10^{-14}\;M_\odot\;yr^{-1}$. This independency of the mass loss rate on level of activity can be explained by the fact that the activity level is governed by the large modulations in the solar close magnetic flux, while the mass loss rate is governed by the rather constant open magnetic flux. I derive a simple expression for stellar mass loss rates as a function of the stellar ambient weak magnetic field, the stellar radius, the stellar escape velocity, and the average height of the Alfv\'en surface. This expression predicts stellar mass loss rates of $10^{-15}-10^{-12}\;M_\odot\;yr^{-1}$ for Sun-like stars.

\end{abstract}



\section{Introduction}
\label{sec:Intro}

\cite{Parker58} introduced the concept of a radially uniform supersonic flow, continuously flowing from an iso-thermal hot corona 
as a result of a hydrodynamic thermal acceleration. One can imagine space as a huge vacuum cleaner that sucks up 
solar material, while solar gravity opposes this force. The gas slowly accelerates sub-sonically until it reaches 
a critical point, where its thermal energy equals the gravitational energy of the Sun. Above this critical point, the 
flow becomes super-sonic and continues to accelerate to some asymptotic terminal speed, $u_\infty$. 
This idea was controversial at first, but have been widely accepted since the first sporadic \citep{Gringauz60} and consistent 
\citep{Neugebauer66} in-situ measurements of the solar wind were made, proving the existence of such flow. 
Parker's model has served as the baseline to all models for the solar wind and stellar winds; 
the main difference between the models 
is the source of the energy accelerating the wind. These models include non iso-thermal winds, polytropic winds, sound 
waves driven winds, dust driven winds, and line-drive winds 
\citep[see][with references therein]{lamerscassinelli99}. 

Parker's model is hydrodynamic: the flow is assumed to be parallel to the magnetic field, and interaction between the flow and the fields is neglected. 
However, stellar magnetic fields play a significant role in the physics of stellar winds since they govern the transport of the ionized gas in the corona 
\citep{lamerscassinelli99}. \cite{Parker58} did not 
include the solar magnetic field in his wind model, but he discussed extensively the consequences of the coupling between the wind and the field 
to conclude the structure of the magnetic field in the interplanetary space. \cite{weberdavis67} have shown 
that a star can lose angular momentum due to the fact that open magnetic field lines in stellar coronae (along where the 
wind is flowing) are attached to the star at one end, but are free at the other end beyond the Alfv\'en point, $r_A$, 
at which the flow speed equals the Alfv\'en speed, $u_A=B/\sqrt{4\pi\rho}$. Here $B$ is the magnetic field strength and $\rho$ is 
the plasma mass density. Coronal plasma is co-rotating with the star, 
but it is also frozen in to the magnetic field. Therefore, a torque is applied on the star by an arm with a length of $r_A$. 
This ``magnetic braking'' explains how stars spin down, where 
the angular momentum loss rate depends on the stellar mass loss rate. Therefore, stellar mass loss rates and 
angular momentum loss rates are crucial inputs for stellar evolution models, which then provide input to other models.

Unfortunately, neither stellar winds nor stellar mass loss rates can be directly observed for low-mass stars
\citep[see][and references therein]{gudel07}. Some attempts to estimate stellar 
mass loss rates were done using chemical separation and H$\alpha$ profiles \citep{Michaud86,LanzCatala92,Bertin95}, 
radio observations \citep{Abbott80,Cohen82,Hollis85,Lim96,Gaidos00}, and observations of x-ray emission due to charge exchange 
\citep{Wargelin01}. 
These calculations have estimated stellar mass loss rates between $10^{-14}-10^{-10}\;M_\odot\;yr$. 

Perhaps the most extensive estimation of stellar mass loss rates was done using measurements of Ly$\alpha$ absorptions 
at the edge of stellar heliospheres, where the stellar wind collides with the Interstellar Medium (ISM) \citep{Wood02,Wood05,Wood06}. 
The formation of a shock at this location causes the build up of a Hydrogen ``wall'' that can be measured in the Ly$\alpha$ 
line. By estimating the ISM properties, the stellar mass flux carried by 
the wind can be determined. Based on many observations of different stellar types, \cite{Wood05} derived a scaling law for 
stellar mass loss rate as a function of x-ray luminosity, which can also be considered as a function of stellar age and activity 
level. They found that young, active stars (such as the young Sun) have mass loss rate of about $10^{-12}\;M_\odot\;Yr^{-1}$, 
which is $\sim$100 times higher than the approximated solar mass loss rate of $2\cdot10^{-14}\;M_\odot\;Yr^{-1}$, derived from typical solar 
wind parameters near Earth. 

The results in \cite{Wood05} are presented cautiously, and the assumptions made are clearly stated. However, 
their intuitive scaling law, which states that the more active the star, the higher the mass flux, is commonly used in an 
{\it ad-hoc} manner, despite of the fact that it might be overestimated. Other calculations for the young Sun revealed a mass loss rate 
which is only 10 times higher than the current value \citep{HolzwarthJardine07,Cohen09,Sterenborg11}. In fact, 
the scaling law in \citep{Wood05} breaks down for the most active stars. In contrast, \cite{Fisher98} and \cite{Pevtsov03}  
used solar data to derive a scaling law between X-ray spectral radiance and magnetic flux. They found a nearly linear relation 
that holds for many orders of magnitude. Therefore, it is unclear how stellar winds and thus mass flux correlate with 
stellar magnetic activity and observed X-ray flux.
 
Unlike the sparse information about stellar winds, the solar wind has been widely observed both remotely and in-situ since the 
beginning of the space era. These observations, together with extended monitoring of the solar magnetic field, have revealed a 
much more complex physics, with the solar magnetic field governing 
the power, the acceleration, and the topology of the solar wind. Magnetic energization of the 
wind is necessary to explain the discrepancy between the fastest wind predicted by Parker's model (about $600\;km\;s^{-1}$) 
and the observed fast solar wind (about $800-900\;km\;s^{-1}$). The exact mechanism in which the magnetic energy is converted to kinetic energy is 
not fully understood yet. Models to explain this energy conversion consider the effect of wave dumping due to electron/ion resonance 
frequencies, dumping of Alfv\'en waves, magnetic energy turbulent cascade, and energy dissipation due to magnetic reconnection 
\citep[see][for detailed description and references therein]{Aschwanden05,Cranmer10}. These models are tightly related to the 
solar corona heating problem. Great progress is currently being made in solving the coronal heating and solar wind acceleration 
problems due to new observations from the HINODE\footnote
{{\tt http://www.nasa.gov/mission\_pages/hinode/index.html}} and SDO\footnote
{{\tt http://sdo.gsfc.nasa.gov}} missions. 

Based on the long-term in-situ measurements of the solar wind taken at 0.3~AU (HELIOS), the outer solar system (Voyager), 
high heliographic latitudes (Ulysses), as well as near Earth (WIND, ACE), some non-trivial properties of the solar 
wind have been revealed. These properties are well summarized in \cite{McComas07} (with references therein). 
A most notable feature is the bi-modal structure of the solar wind, which seems to be composed of two different populations. 
The first population is the so-called ``fast wind''. It originates from coronal holes --  
the large-scale regions of open field lines (polar regions during solar minimum) -- and it is rather steady, 
with speeds of the order of $750\;km\;s^{-1}$ or more. The second population is the so-called ``slow wind''. 
This wind is much more sporadic than the fast wind and it is assumed to originate from the vicinity of the large, low latitude helmet 
streamers during solar minimum. During solar maximum, the slow wind is observed to originate from higher latitudes, as well as 
from the vicinity of active regions. The two populations are not distinguished by flow speed only, but also by densities, temperatures, 
composition, abundances, distribution functions, and First Ionization Potential (FIP) levels. While the acceleration of the fast 
solar wind could, in principle, follow Parker's model, with the contribution of magnetic energy by one of the proposed mechanisms above, 
the slow solar wind introduces a greater challenge in explaining its acceleration mechanism. An observed inverse 
correlation between the flow speed and electron temperature, and a strong ion temperature anisotropy with very hot minor species at 
the low corona are inconsistent with the theory of thermal acceleration. In general, it seems that both populations are at first confined 
in closed loops, then are later released along open field lines. The fast wind is initially confined near coronal holes, where loops are 
small, cool, less dense, and short-lived, while the slow wind is confined within the large, hot, dense loops, with the plasma spending more 
time in the loop before it is released \citep{Feldman99}. \cite{Fisk99} showed how the mass flux and Poynting flux at the 
flux-tube base together with the lifetime of the loops are sufficient to explain the fast solar wind and determining 
its final speed. 

The difference in final speed suggests that the two populations gain different amounts of energy through the acceleration process. 
This can be generally explained by the different energy per unit mass the populations have (due to their different densities). 
In addition, due to its fast release the fast wind has more time to gain electromagnetic energy compared to the slow wind. 
The frequent ejection of the fast wind also explains why it is quasi-steady compared to the sporadic slow wind. Alternatively, 
the two populations can gain the same amount of energy at the coronal base but have different amount of energy losses in the acceleration process. 
\cite{SchwadronMccomas03} proposed that both populations gain similar electromagnetic energy to have high speed values 
($\sim 800\;km\;s^{-1}$). The difference in the final kinetic energy results from the wind radiating
energy in the form of heat conduction and the advection of thermal energy as it is accelerated, where these radiative terms mostly depend 
on the maximum temperature of the loops. The loops can be thought as a reservoir of energy that is being transferred to the wind once 
it is released along open field lines. Since the slow wind is released from large, long-lived, hot loops, it radiates much more 
then the fast wind, so that its final kinetic energy is smaller. They also showed that their scaling law and its dependency on 
the maximum loop temperature explains the observed anti-correlation between the electron temperature and final wind speed, 
the FIP effect \citep{Schwadron99}, and the anti-correlation between the magnetic flux-tube expansion factor and the 
wind speed \citep{wangsheeley90}.     

\cite{SchwadronMccomas06}, and \cite{SchwadronMccomas08} further showed that the energy function presented in \cite{SchwadronMccomas03} 
can be modified to provide a linear correlation between the solar X-ray power and the {\it open} magnetic flux at the base of the flux-tube. 
They showed that solar wind data obey this linear relation, that it can be extended to many orders of magnitude, and that it is similar to the 
relation presented in \cite{Pevtsov03}. Recently, \cite{Wang10} have estimated the mass and energy flux at the coronal base using solar wind 
data taken by ACE and Ulysses in combination with solar magnetic field (magnetogram) data. Based on conservation laws along a magnetic flux 
tube, he derived an identical formula as in \cite{SchwadronMccomas06}. The only difference is that while \cite{SchwadronMccomas06} assumed a 
constant value for the wind speed far from the Sun, \cite{Wang10} has used actual data for this parameter. He found that most of the energy 
flux at the coronal base is used to lift the wind from the gravitational well, i.e. provides the wind with the escape velocity. He also found that while 
the magnetic field, the mass flux, and energy flux vary a lot at the coronal base, these parameters are rather constant 
at 1~AU where the variations are balanced by the geometrical expansion of the magnetic flux tube. Similarly, \cite{SchwadronMccomas06} explained 
why the solar X-ray luminosity vary by an order of magnitude, while the observed wind power is rather constant. This is due to the fact that the 
strong solar X-ray emissions come from the closed, hot loops and the wind power is dominated by the open, rather constant solar 
magnetic flux \citep{Wang00} . Both \cite{SchwadronMccomas06} and \cite{Wang10} have suggested  that the solar wind fluxes are determined by 
the large-scale, dipolar component of the solar magnetic field, as the 
heliosphere indeed seems to be composed by two hemispheres with opposite polarity separated by a single current sheet, which is reversed through 
the solar cycle \citep{Smith01,Jones03}.

In this paper, I argue that the X-ray flux observed on stars (or stellar activity in general) cannot be a reliable proxy for its wind mass flux and 
power. In Section~\ref{SWdata}, I investigate the dependency of the solar wind mass loss rate on the solar X-ray flux using solar wind 
data taken at different periods of time and at different locations in the solar system. In Section~\ref{sec:Discussion}, I discuss these results 
and derive a relation between stellar mass loss rate, and the stellar magnetic and gravitational parameters. I conclude my findings in 
Section~\ref{sec:Conclusions}.


\section{Observations of Solar Wind Mass Loss Rate and Solar X-ray Flux}
\label{SWdata}

In this section, I study the relation between the solar wind mass loss rate and the solar X-ray flux, which is the proxy commonly used to relate 
stellar mass loss rates and the level of stellar activity. I use solar wind in-situ measurements taken by WIND/ACE (near 1~AU), 
and by Ulysses (at high heliographic latitudes between 3-5~AU), 
as well as measurements of solar soft X-ray (1-8~$\AA$) flux taken by GOES 7 and 8\footnote
{Obtained via NOAA's GOES, at {\tt http://goes.ngdc.noaa.gov/data}}, during solar cycle 23 (1996-2006). 
To study longer-term trends, I use solar wind observations for the years 1965-2010\footnote
{Obtained via NASA's OMNIWEB, at {\tt http://omniweb.gsfc.nasa.gov}}, Voyager II observations of the solar wind at the outer solar system, 
and long-term sunspot data\footnote
{Obtained via the Solar Influance Data Analysis Center, at {\tt http://www.sidc.be/sunspot-data}} \citep{sidc9606}. ACE, WIND, Voyager, 
and Ulysses data have all been obtained from the CDAWEB website\footnote
{Obtained via NASA's CDAWEB, at {\tt http://cdaweb.gsfc.nasa.gov}}.

\subsection{Data Selection}
\label{sec:DataSel}

Here, we are interested in comparing the {\it global, ambient} solar mass loss rate with a particular 
solar activity level. There are three challenges in this context when using in-situ data. First, transient events, such as 
Coronal Mass Ejections (CMEs) and Corotating Interaction Regions (CIRs) can appear in the data, while they do not represent the 
ambient solar wind. Second, solar flares and short-term significant increases in the X-ray data also do not represent the ambient 
X-ray flux. Third, an in-situ measurement of the solar wind does not necessarily represent the global value of some parameter. In order 
to avoid these issues, I select the solar wind data used in this study as follows. 

I first concentrate on the data available for solar cycle 23 (taken between 1996-2006). I use hourly averaged data to obtain a 
large amount of data points. I exclude however, any impulsive, short-lived spikes 
so that the statistical behavior is dominated by the ambient values. Using the ICME list by \cite{CaneRichardson03}\footnote
{Complete list up to 2006, at {\tt http://www.ssg.sr.unh.edu/mag/ace/ACElists/ICMEtable.html}}, I exclude periods which are 
associated with Interplanetary Coronal Mass Ejections (ICMEs) measured at 1~AU from the data. 
For Ulysses data, there is no such list. Therefore, I remove data periods when both the number density is higher than $10\;cm^{-3}$ 
and the magnetic field magnitude is higher than $10\;nT$. Even though this method is robust, it is most unlikely to have such conditions 
in the ambient solar wind observed by Ulysses. 

The GOES data is provided in units of $Watts/m^2$ as observed at 1~AU. The X-ray flux units commonly used in the astronomy literature are 
$ergs/(cm^2s)$, where the X-ray luminosity is divided by the stellar surface. In order to be consistent with these units, I converted the GOES flux in to 
cgs units, and also used the factor $(1AU)^2/R^2_\odot$ to obtain the X-ray flux units in the astronomical context.  
When using the X-ray flux as the global representation for the solar level of activity, solar flares and short-lived data peaks must be removed. 
In order to exclude solar flares from the GOES data, I remove data points with X-ray flux higher than $45\;ergs\;cm^{-2}\;s^{-1}$ 
\citep[$10^{-6}\;W/m^2$ - the typical flux associated with flares is above this value, see][]{Golub09}. Typical non-flaring X-ray fluxes in the GOES data 
are of the order of $0.1-30\;ergs\;cm^{-2}\;s^{-1}$ ($5\cdot10^{-9}-10^{-7}\;W/m^2$), which are three orders of magnitude lower than the total solar X-ray flux - 
$F_{x}\approx 2\cdot10^4\;ergs\;cm^{-2}\;s^{-1}$ \citep{Wood06,Golub09}. 

\subsection{Results}
\label{sec:Res}

\begin{figure*}
\centering
\includegraphics[width=6.4in]{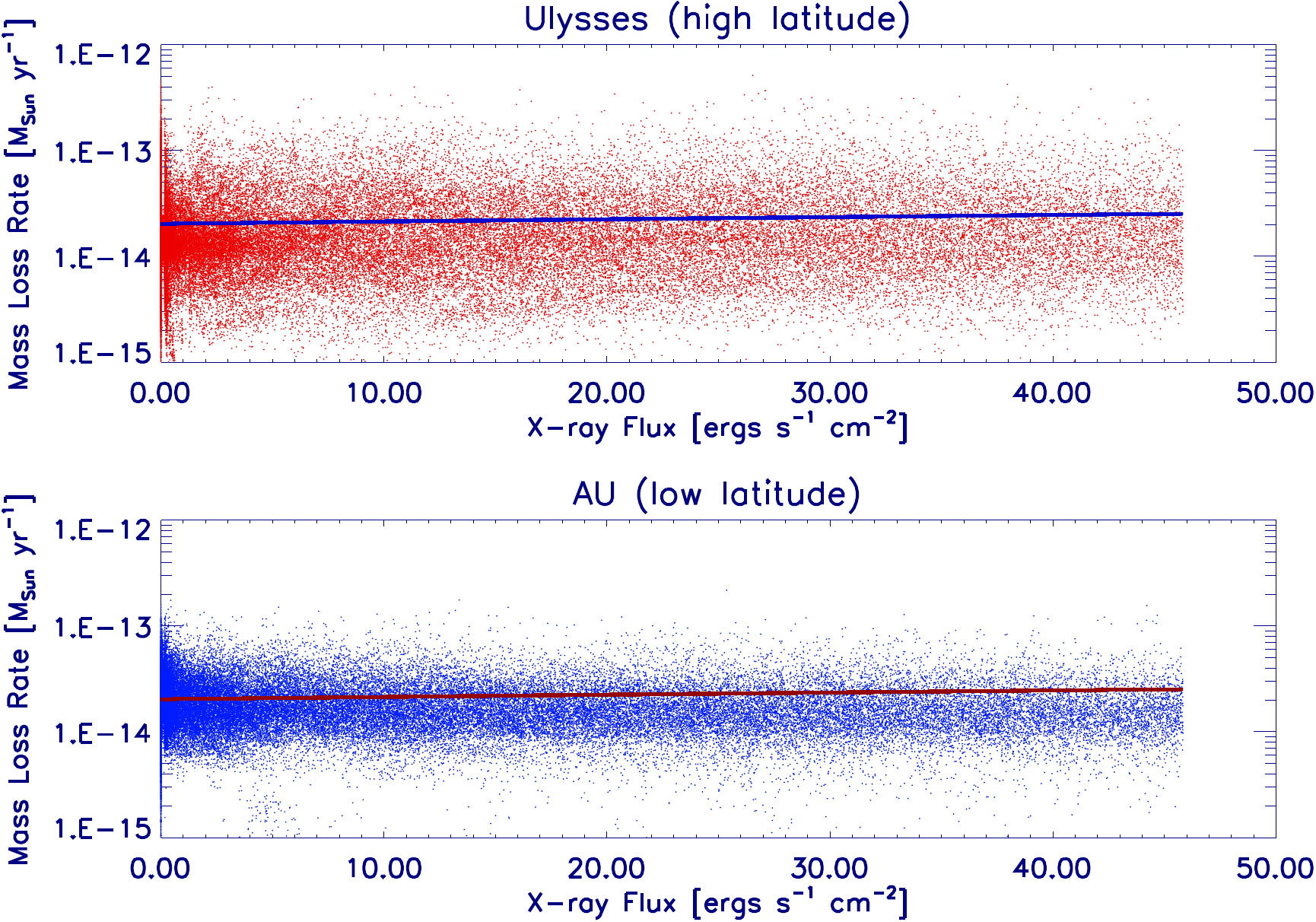}
\caption{Solar mass loss rate as a function of X-ray flux for {\it all} Ulysses data (top) and observation near Earth (bottom) taken 
between 1996-2006.}
\label{fig:f1}
\end{figure*}

Figure~\ref{fig:f1} shows scatter plots of the solar mass loss rate as a function of the solar X-ray flux for {\it all} valid data points 
measured near Earth and by Ulysses between 1996-2006. It can be clearly seen that the best linear fits have flat slopes, and while the 
X-ray flux varies by an order of magnitude, the mass loss rate is uniformly concentrated around the value of 
$\dot{M}_\odot\approx 2\cdot 10^{-14}\;M_\odot\;yr^{-1}$ for all X-ray flux values. The scatter around this constant value is slightly larger 
in the Ulysses data, probably due to the fact that Ulysses have measured more heterogeneous wind (fast and slow), while the 1~AU data contains mostly 
slow wind measurements. 

The best way to determine the global solar mass loss rate from in-situ data, is to combine simultaneous measurements 
taken at different positions. 
In particular, it is useful to use the available multi-point solar wind data taken at the same time by spacecraft near Earth at low 
heliographic latitude, and by Ulysses at high heliographic latitude. Fortunately, Ulysses was at high latitudes during 1996 (solar minimum) 
and 2000 (solar maximum). This enables us to investigate the properties of the slow and fast solar wind separately, while giving a good 
estimate of the total solar mass loss rate at a particular time.  
   
\begin{figure*}
\centering
\includegraphics[width=3.2in]{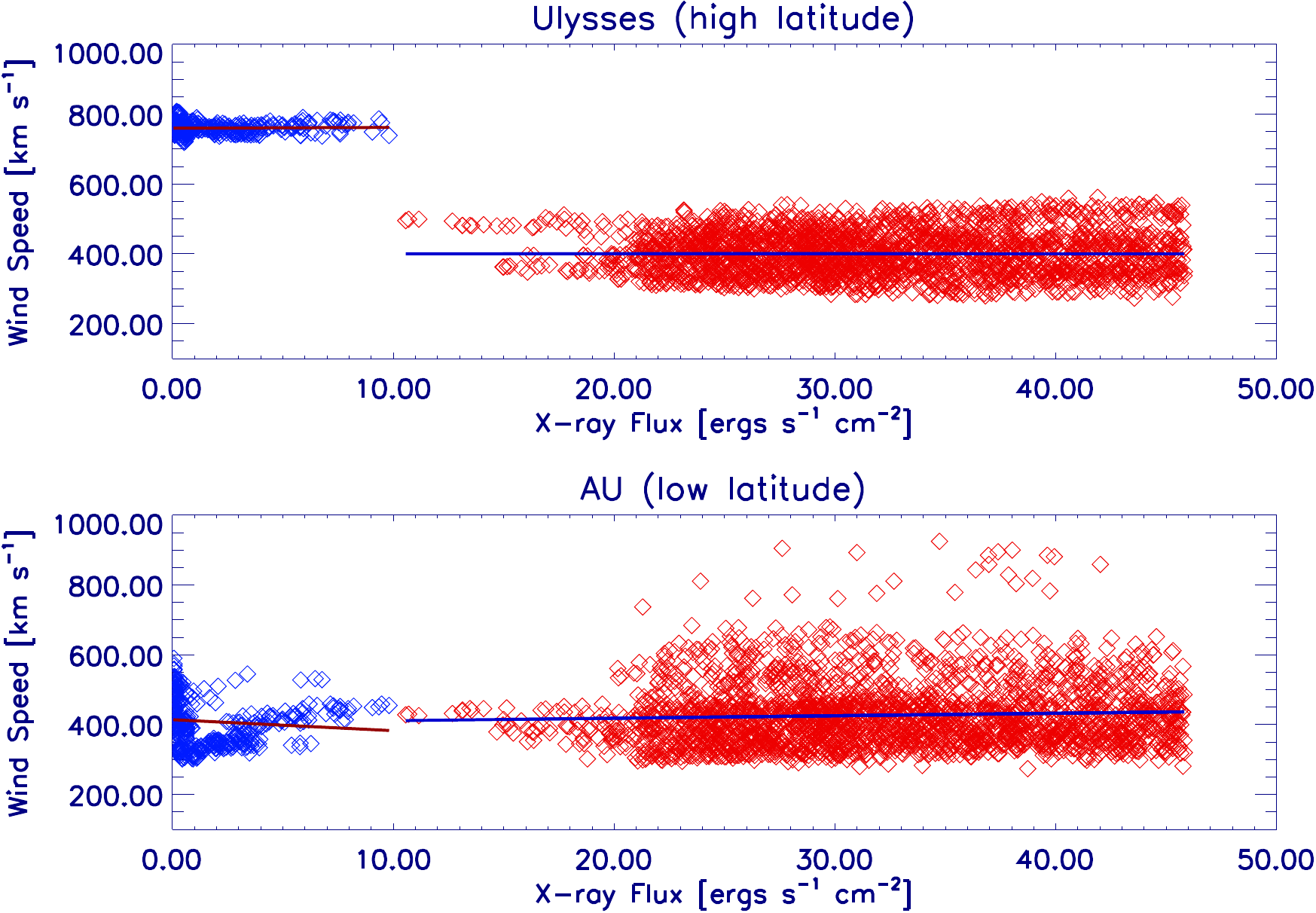}
\includegraphics[width=3.2in]{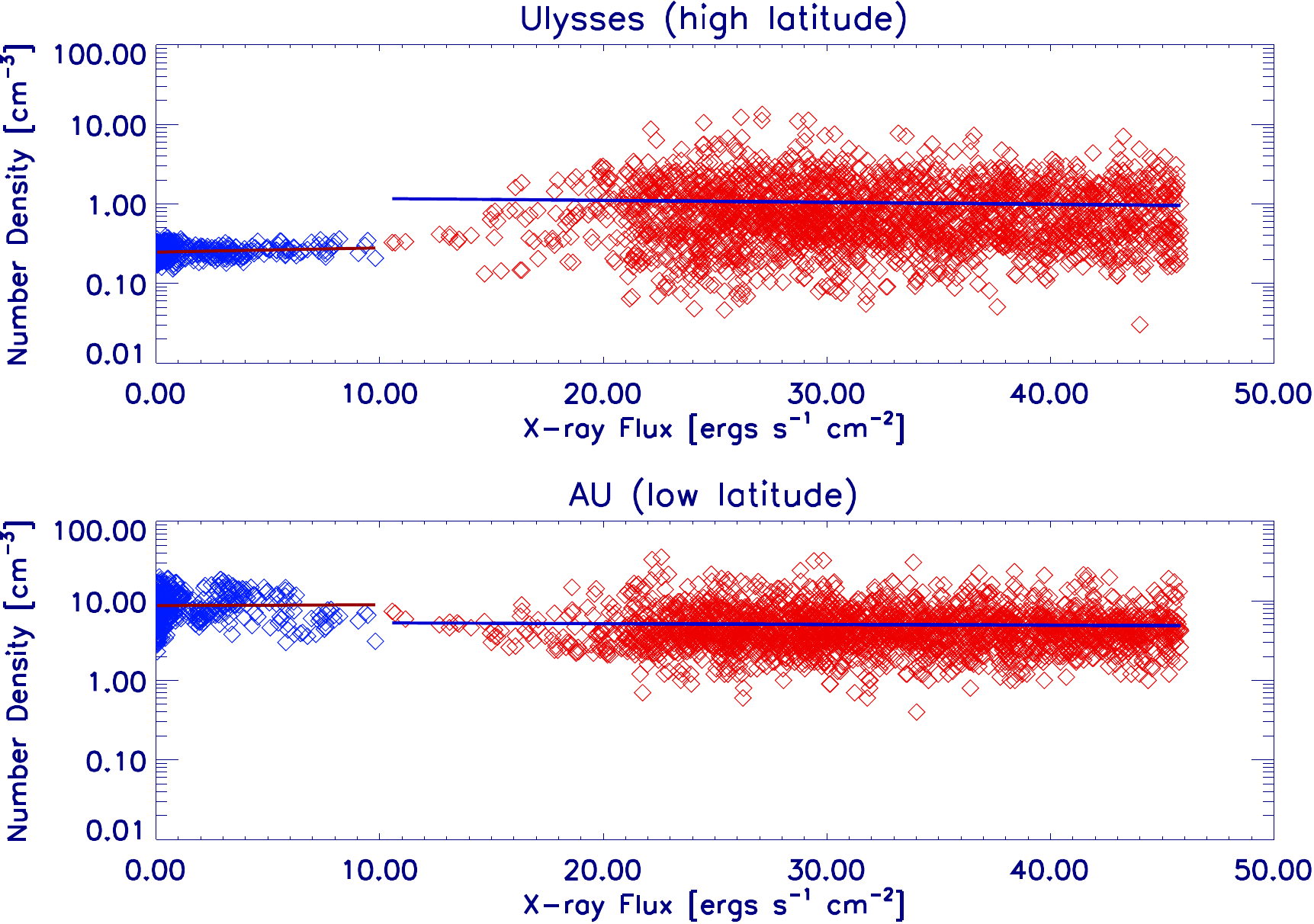}
\caption{Solar wind speed (left) and number density (right) as a function of X-ray for Ulysses data (top) and observation near 
Earth (bottom) taken during solar minimum (1996, blue) and solar maximum (2000, red).}
\label{fig:f2}
\end{figure*}

Figure~\ref{fig:f2} shows scatter plot of the wind speed and number density as a function of the X-ray flux, only for data points taken 
during the year 1996 and the year 2000. Data points for the two years are shown in different colors. Figure~\ref{fig:f2} clearly shows that 
during 1996, Ulysses has measured pure fast solar wind, while during 2000 (solar maximum), it has measured slow solar wind, and 
the interplanetary space was filled with slow wind (even at high latitudes). During solar maximum, the large number of strong active regions 
appear on the solar surface take over the large-scale weak background dipole field component. As a result, 
much of the fast wind is eliminated and large amount of the solar wind originates from non-coronal hole regions \citep{Schrijver03,Cohen09}.

\begin{figure*}
\centering
\includegraphics[width=3.2in]{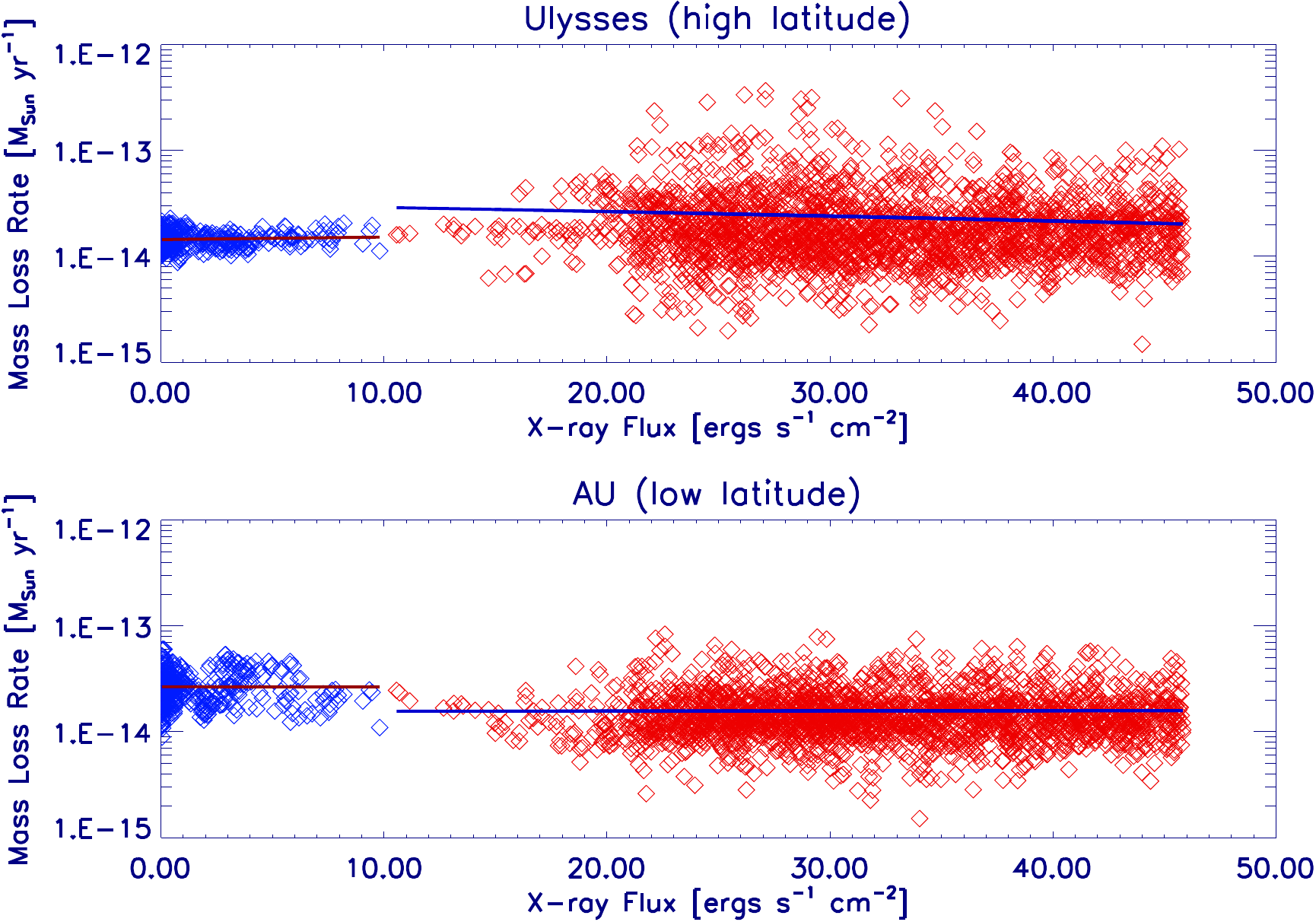}
\includegraphics[width=3.2in]{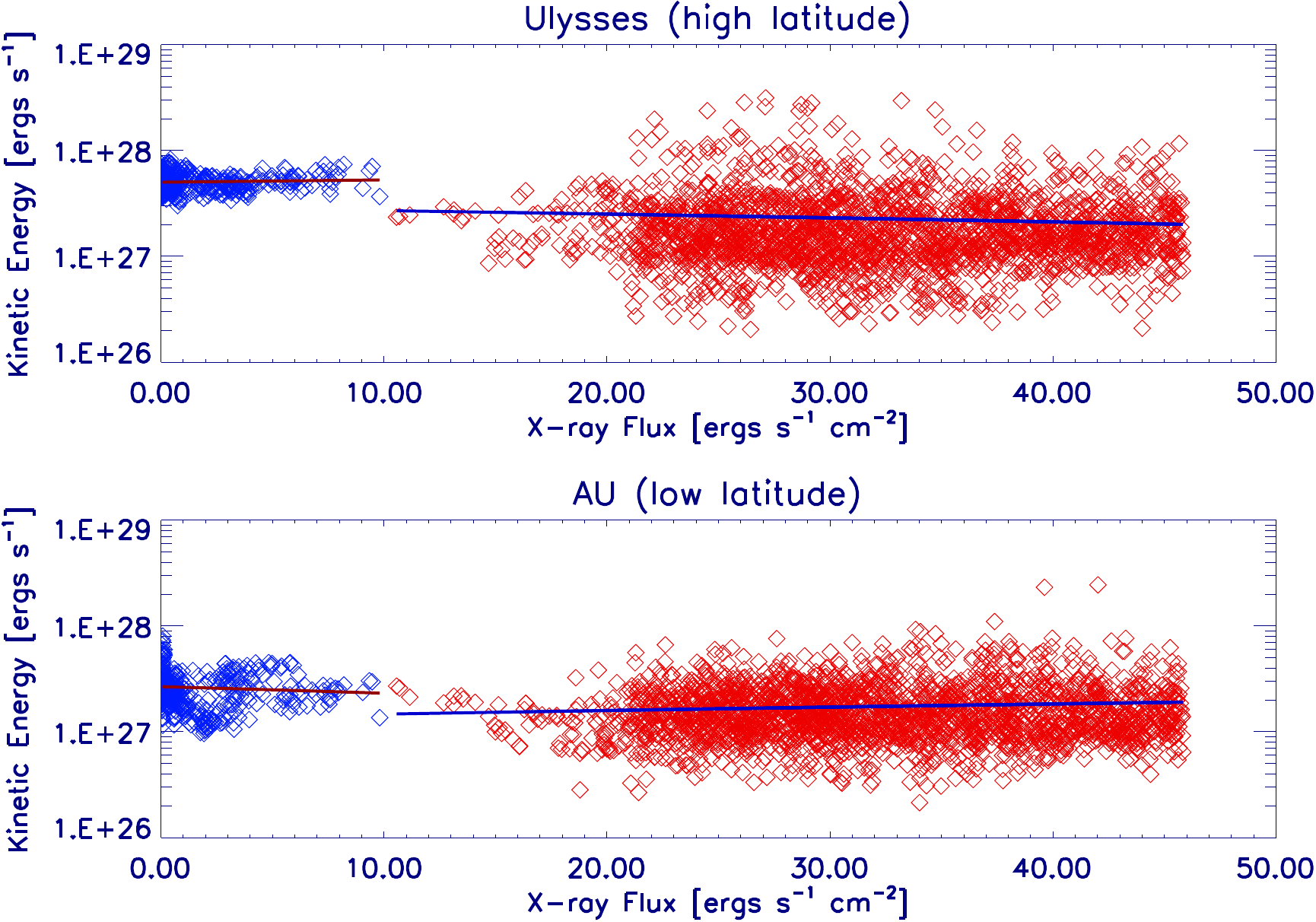}
\includegraphics[width=3.2in]{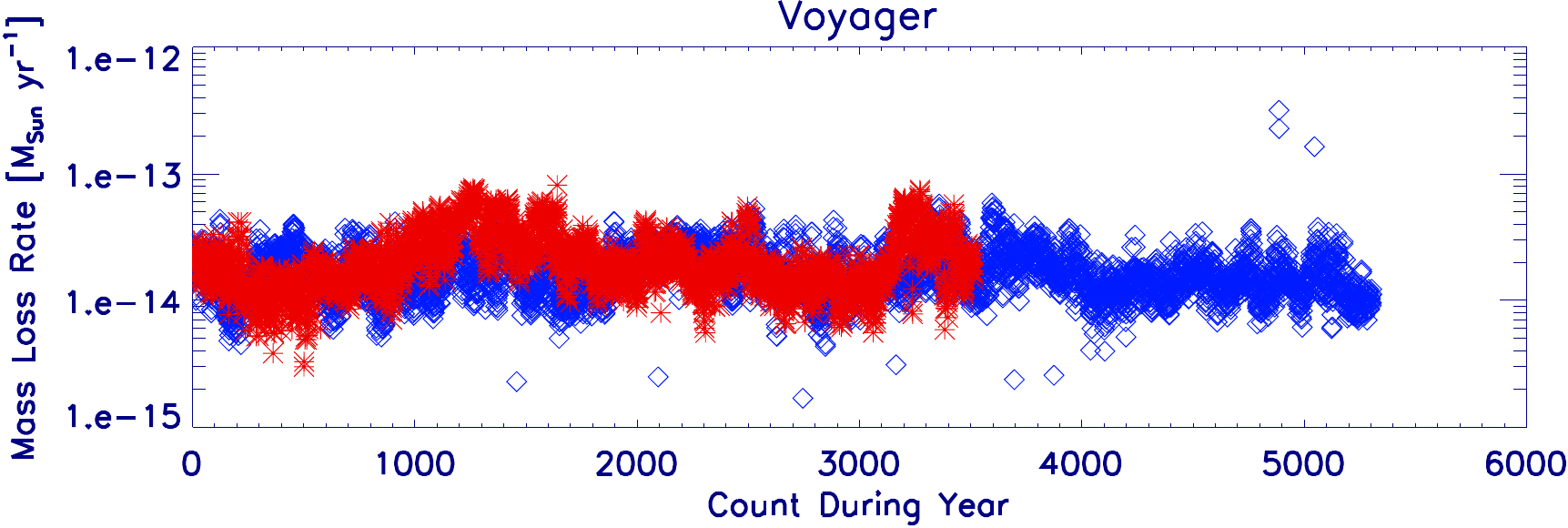}
\includegraphics[width=3.2in]{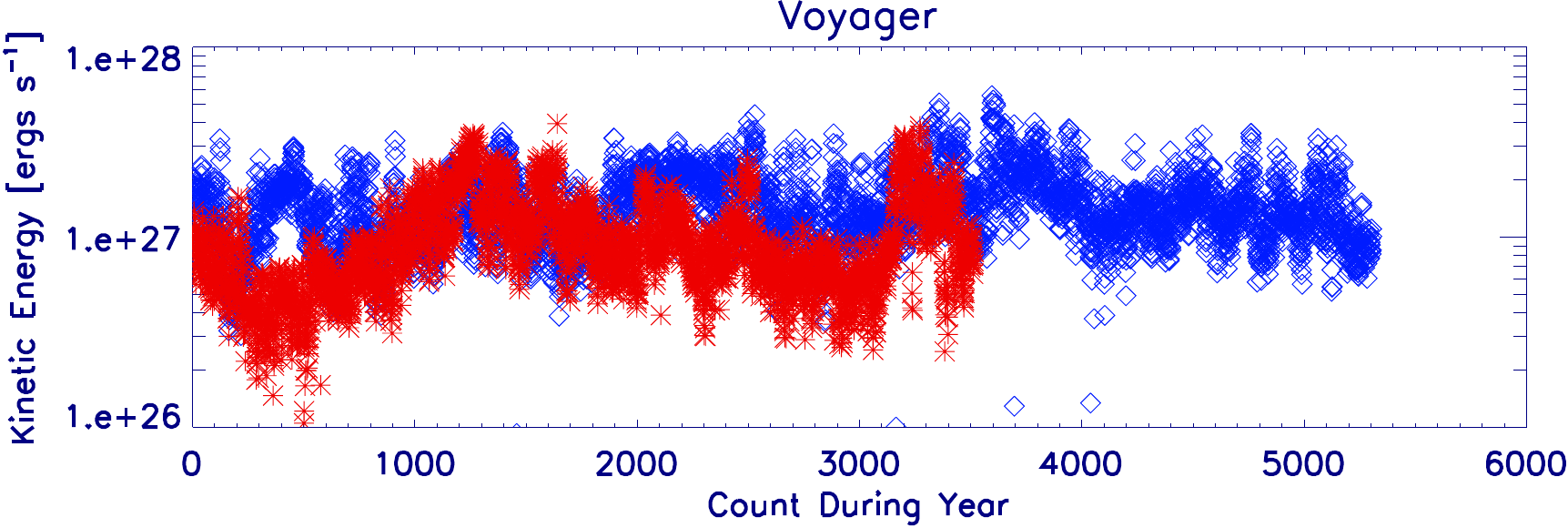}
\caption{Top two panels - solar mass loss rate (left) and solar kinetic energy (right) as a function of X-ray flux for Ulysses data 
(top) and observation near Earth (bottom) taken during solar minimum (1996, blue) and solar maximum (2000, red). Bottom - 
solar mass loss rate (left) and solar kinetic energy (right) as a function of count through the year taken during solar 
minimum (1996, blue) and solar maximum (2000, red).}
\label{fig:f3}
\end{figure*}

In the four upper panels of Figure~\ref{fig:f3}, I show the solar mass loss rate and the solar wind kinetic energy as a function of the solar 
X-ray flux for data taken near 1~AU and by Ulysses during 1996 and 2000 (similar display as in Figure~\ref{fig:f2}). The X-ray flux in 
Figure~\ref{fig:f3}, as well as Figure~\ref{fig:f2}, has a clear separation between solar minimum period, with values 
below $10\;ergs\;s^{-1}\;cm^{-2}$, and solar maximum period, with higher values ranging between $10-45\;ergs\;s^{-1}\;cm^{-2}$. 
The solar mass loss rate and the solar wind kinetic energy are concentrated around $2-4\cdot10^{-14}\;M_\odot\;yr^{-1}$ and 
$5\cdot10^{27}\;ergs\;s^{-1}$, respectively. The solar wind observed during 1996 was very steady and concentrated around these values, 
while the solar wind observed during 2000 is more sporadic, with larger variations around the mean value. Interestingly, the fast wind 
seems to be slightly more energetic than the slow wind, while there seems to be no significant difference in their mass fluxes. 
To provide an additional reference observation, I also plot the solar mass loss rate and the solar wind kinetic energy as a 
function of time based on data taken by Voyager II during 1996 and 2000. At that time, Voyager II was located at the outer part of the 
solar system. Here I simply look at the values of these parameters throughout a year of a quiet Sun and an active Sun, without a direct 
comparison with the X-ray flux. The Voyager data shows variations of the solar mass loss rate, which are similar 
to those plotted in the upper panels of Figure~\ref{fig:f3}, and with a similar spread around the same average value. The solar wind 
kinetic energy is rather steady, but is slightly lower than the energy observed at inner parts of the solar system, in particular at 
the beginning of the year 2000. This can be due to solar wind energy exchange with pickup-ions, CIRs, or other temporal variations 
of the solar wind at that distance, which cannot be identified by this time-series alone.

\begin{figure*}
\centering
\includegraphics[width=3.2in]{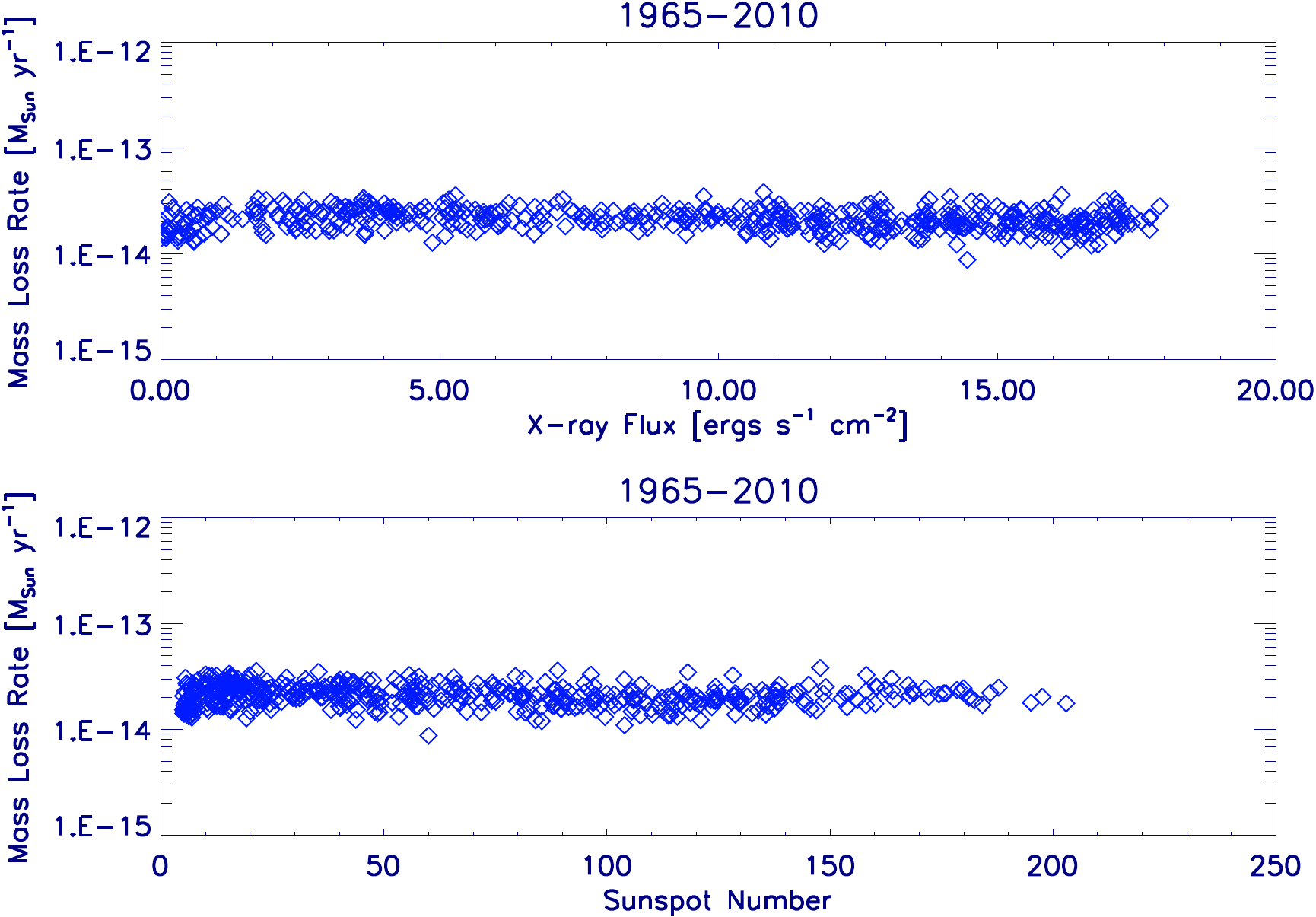}
\includegraphics[width=3.2in]{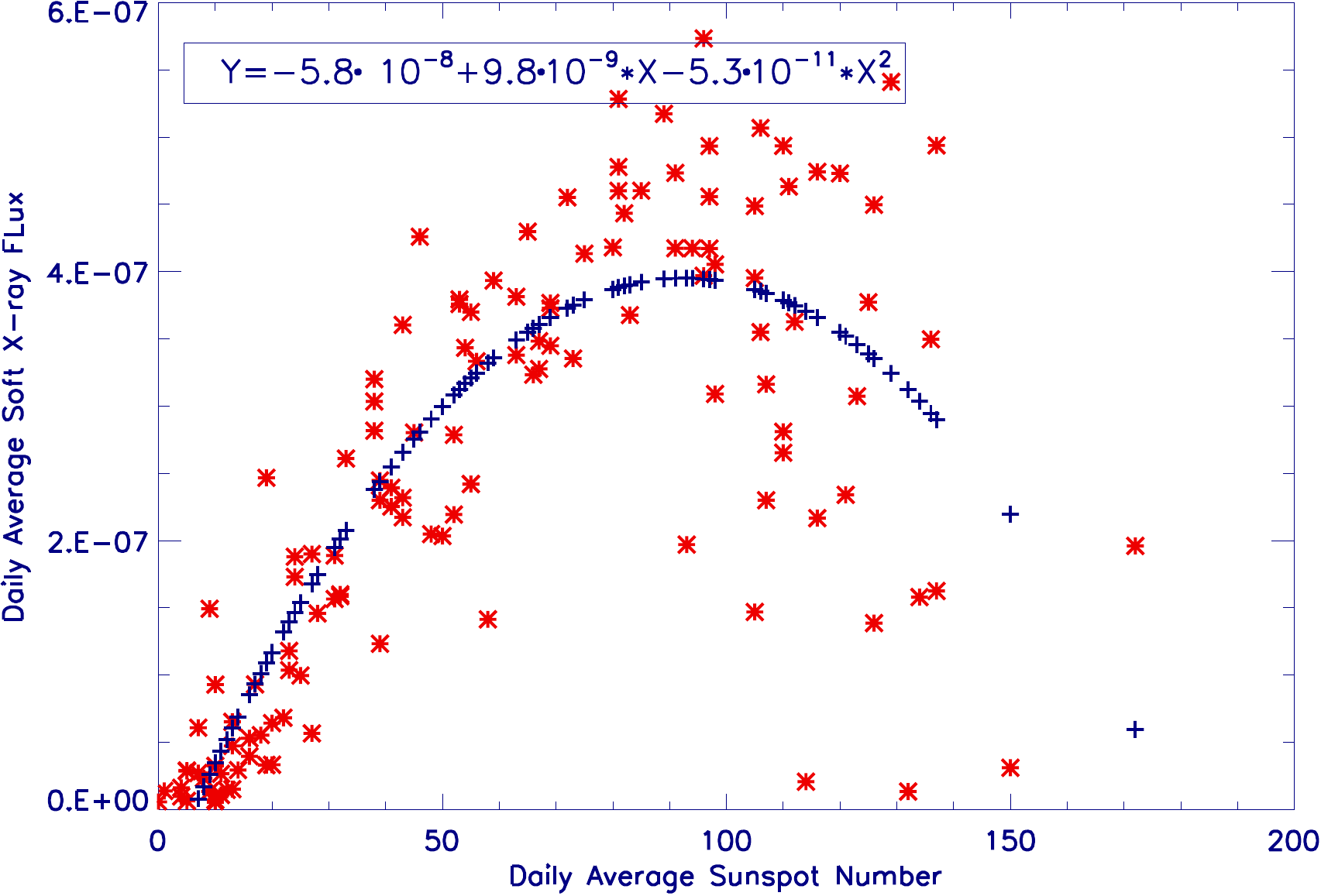}
\includegraphics[width=3.2in]{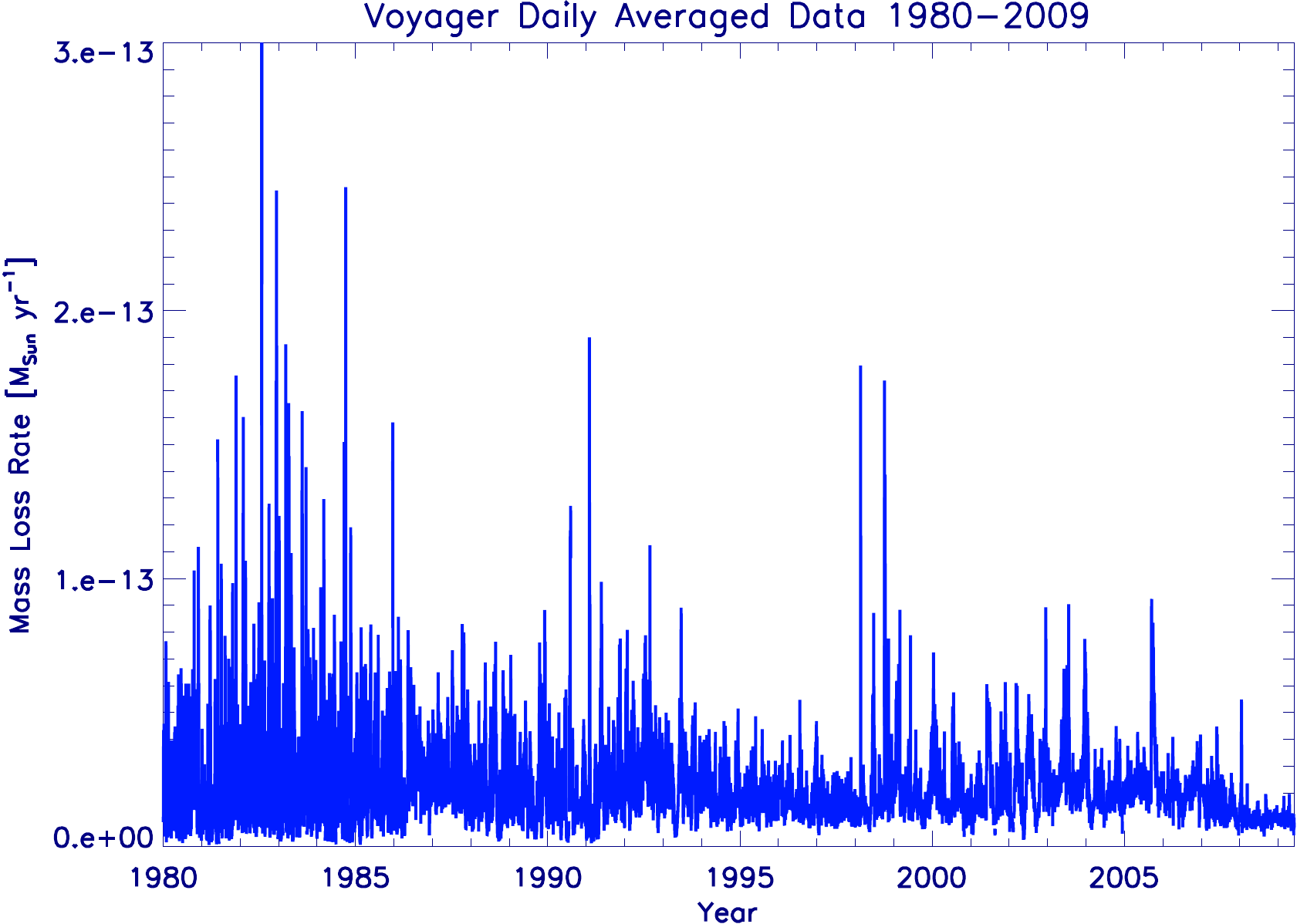}
\includegraphics[width=3.2in]{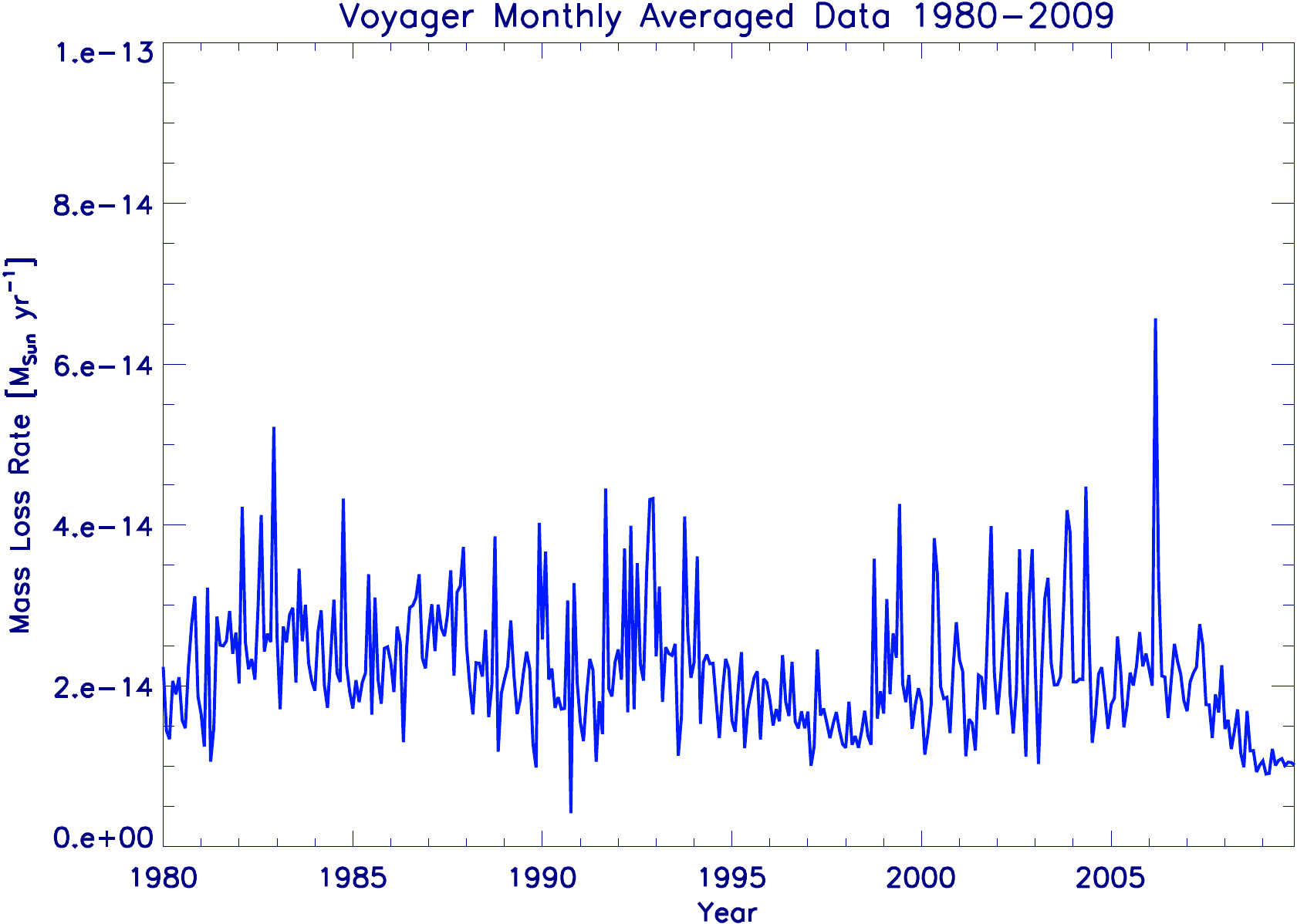}
\caption{Top-left - long-term solar mass loss rate as a function of sunspot number and a scaled X-ray flux for data taken 
between 1965-2010 (monthly averages). Top-right - the scaling function of the X-ray flux with the sunspot number used in 
the top-left plot. Bottom - solar mass loss rate as a function of time for Voyager II data averaged over a day (left) and a month (right).}
\label{fig:f4}
\end{figure*}

A longer-term analysis of the solar wind properties as a function of X-ray flux can be obtain by relating the X-ray flux with the most 
trivial solar activity proxy - the number of sunspots. I use sunspot data for the period when X-ray data is available to obtain a 
simple scaling of the X-ray flux with sunspot number (seen in the upper-right panel of Figure~\ref{fig:f4}). The upper-left panels 
of Figure~\ref{fig:f4} show scatter plots of the monthly averaged solar mass loss rate as a function of both the sunspot number and 
the scaled X-ray flux (similar to Figure~\ref{fig:f1}, Figure~\ref{fig:f2}, and Figure~\ref{fig:f3}) for data taken near Earth 
between 1965-2010. It can be seen that here the slope is even more flat, and that when averaging the solar mass loss rate over a 
month, there is even less scatter around the same value of $2\cdot10^{-14}\;M_\odot\;yr^{-1}$. The two bottom panels of Figure~\ref{fig:f4} 
show similar plots as in the bottom of Figure~\ref{fig:f3}, for daily and monthly averaged Voyager II data taken between 1980-2009. These plots also indicate an average 
solar mass loss rate of $2\cdot10^{-14}\;M_\odot\;yr^{-1}$.

\begin{figure*}
\centering
\includegraphics[width=3.2in]{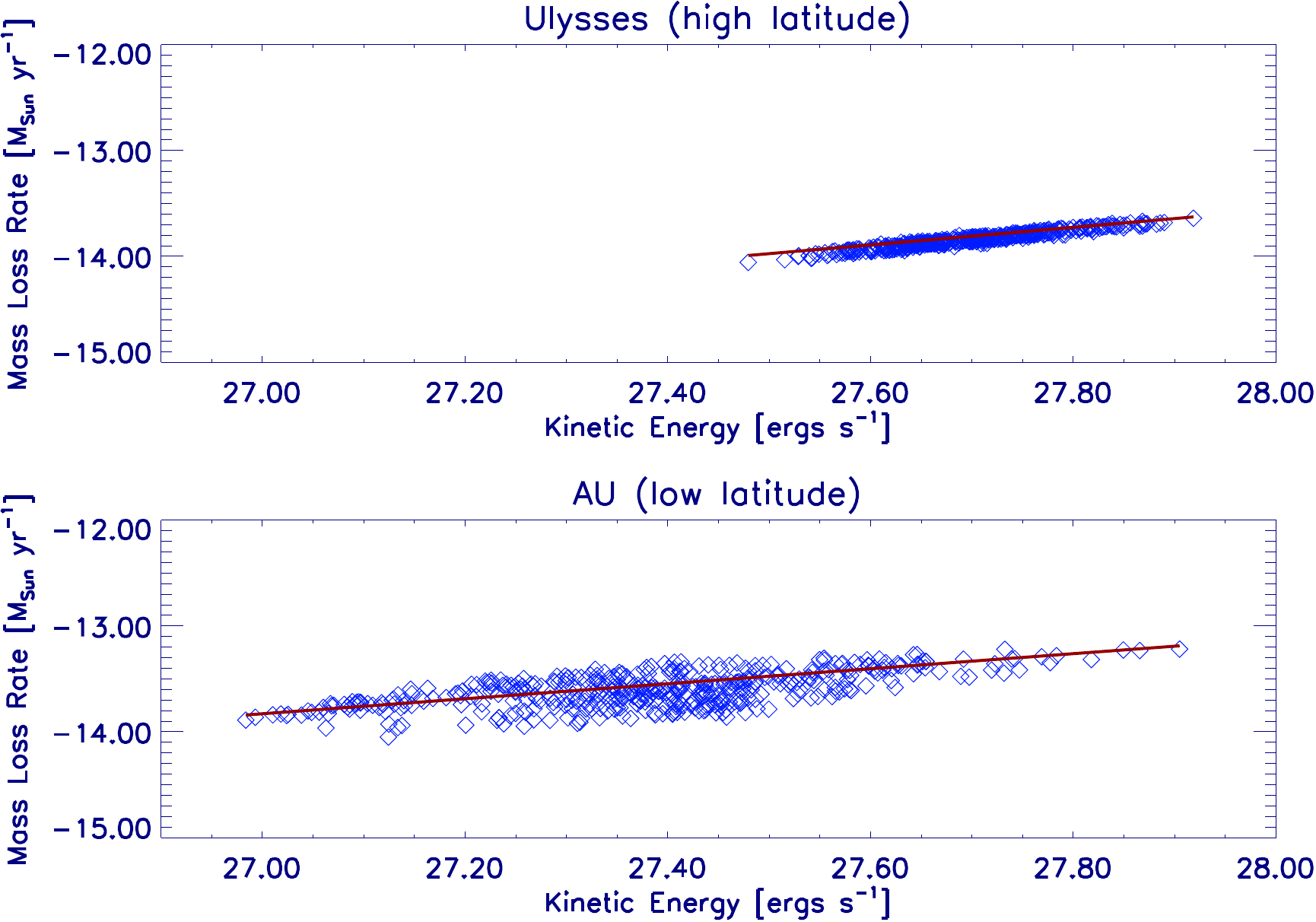}
\includegraphics[width=3.2in]{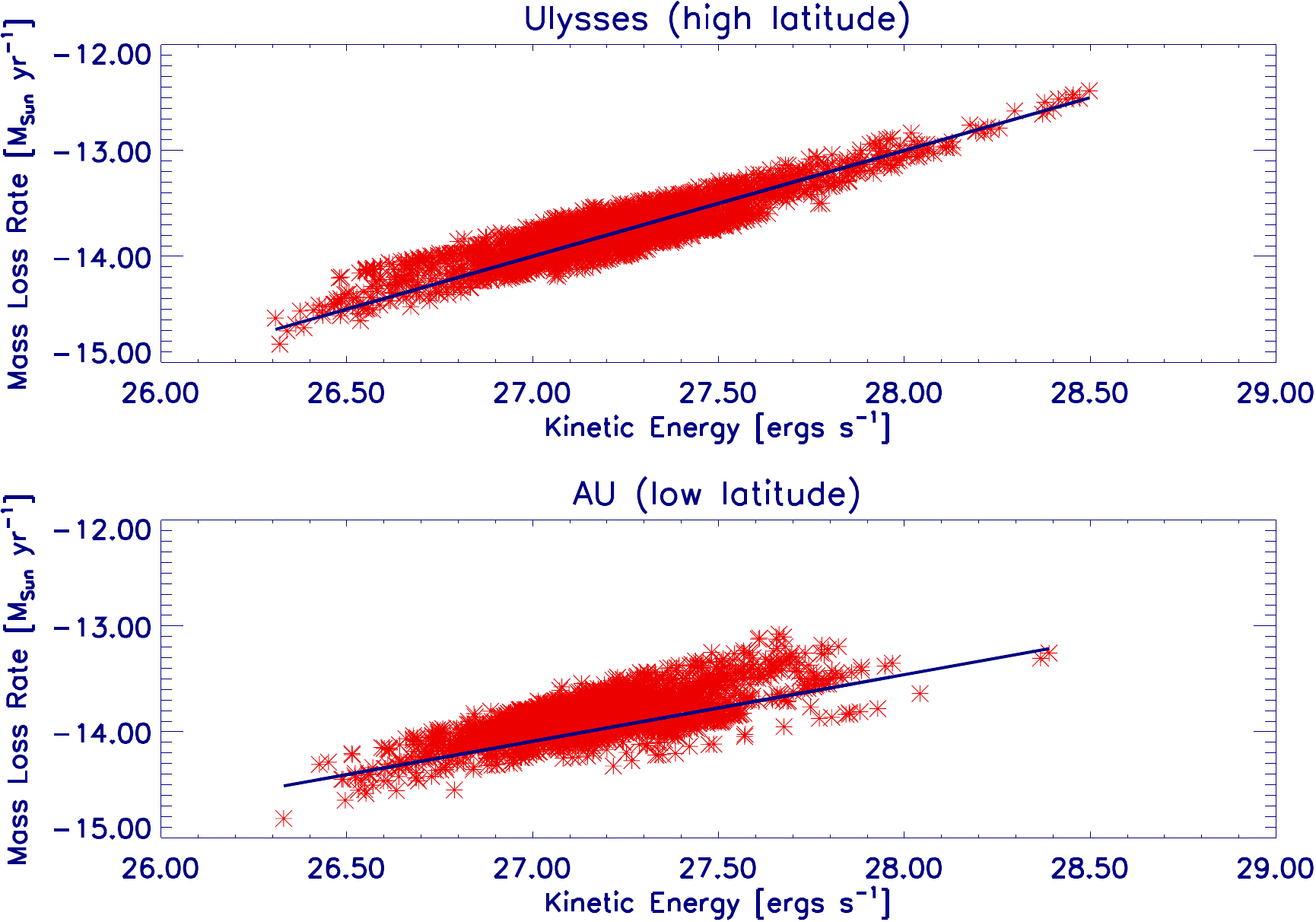}
\includegraphics[width=3.2in]{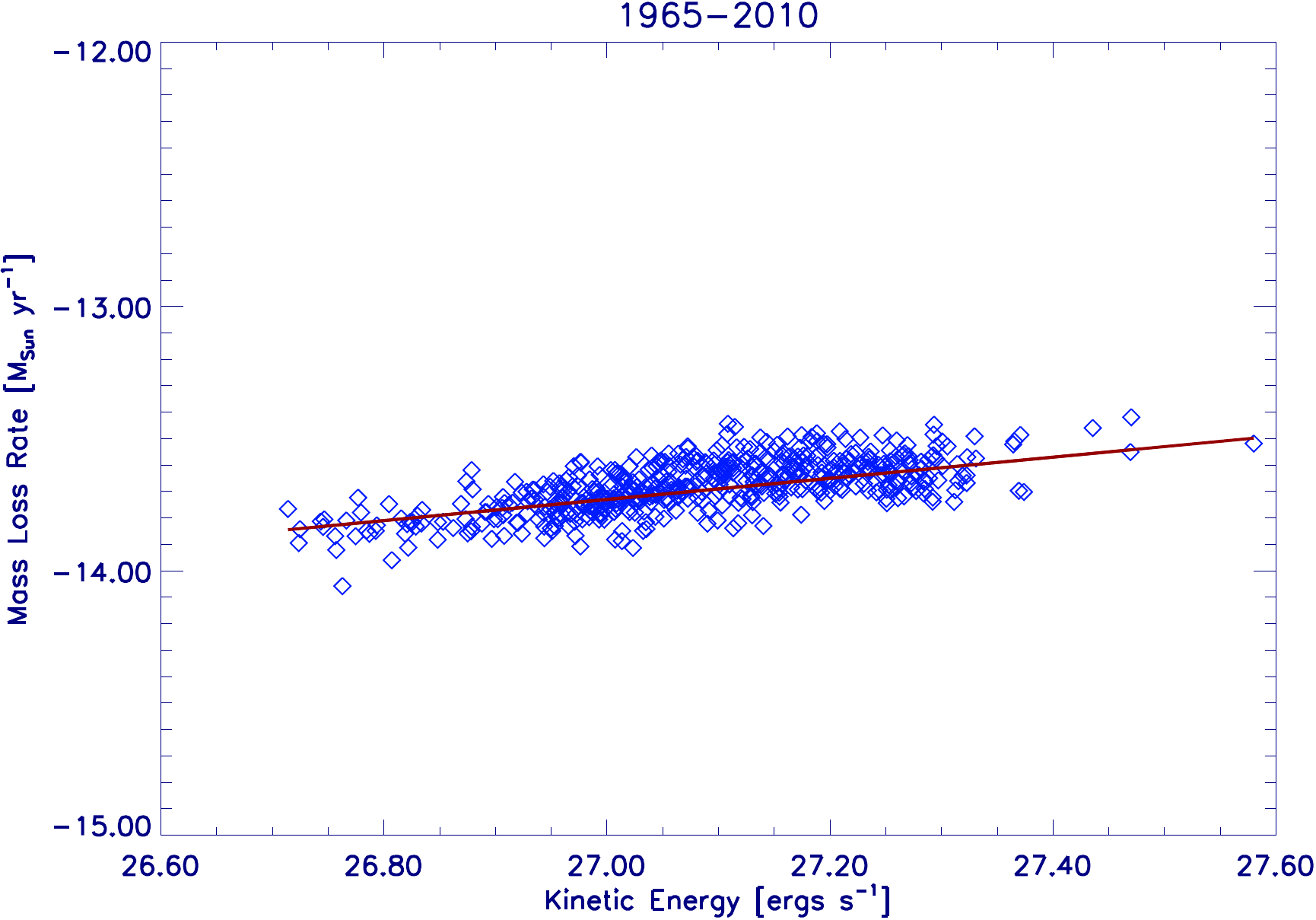}
\includegraphics[width=3.2in]{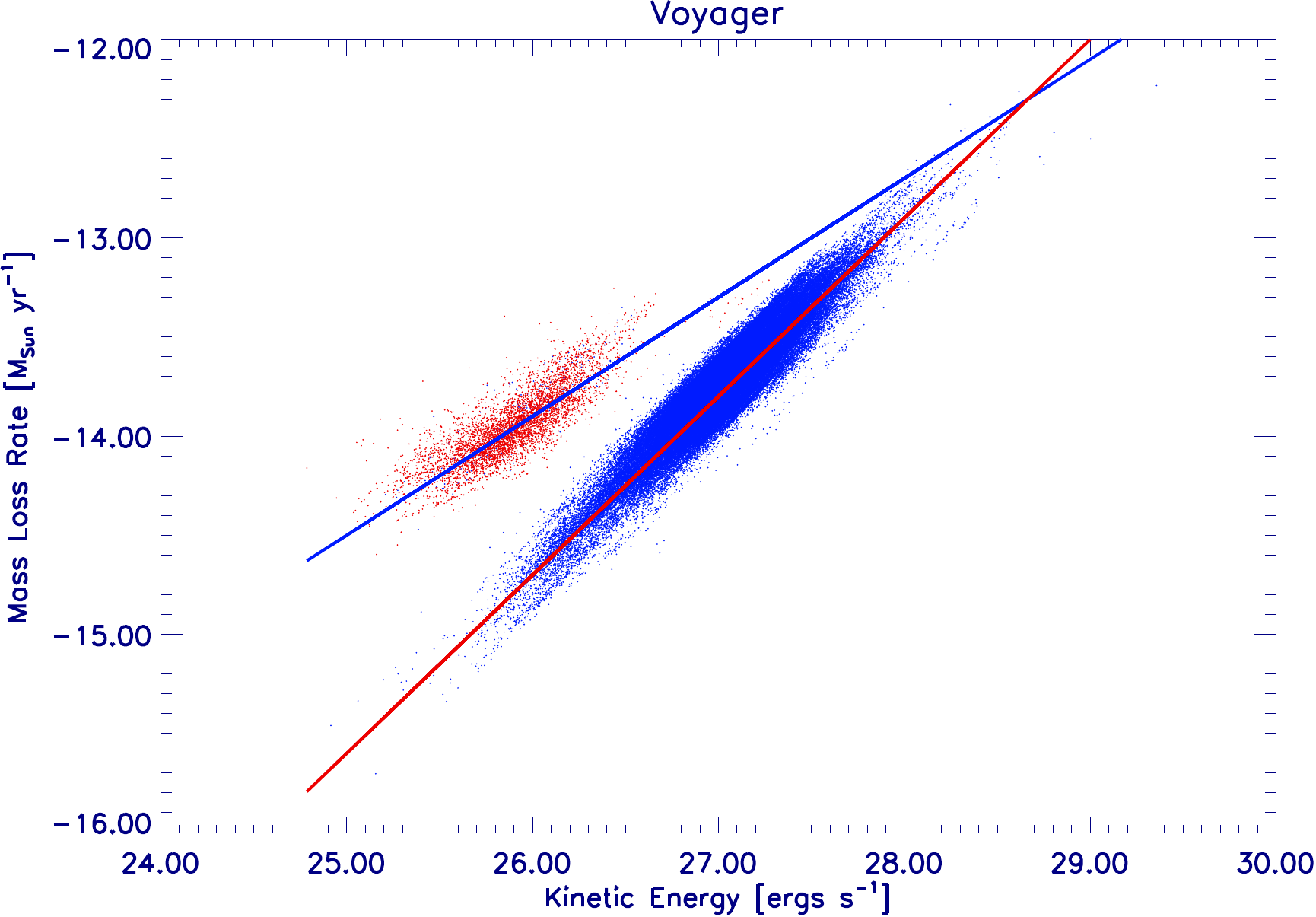}
\caption{Top and middle panels: solar mass loss rate as a function of wind kinetic energy (both in logarithmic scale) for Ulysses (top) and 1~AU (middle) date 
taken during 1996 (left) and 2000 (right). Bottom-left: similar plot for date taken near 1~AU between 1965-2010. Bottom-right: similar plot for data taken 
by Voyager II between 1980-2009. Blue dots are of the solar wind, while red dots are of measurements of the Heliosheath.}
\label{fig:f5}
\end{figure*}

\cite{Wang10} has shown that there is a nearly perfect correlation between the mass and energy fluxes at the base of the corona. 
He stated that this is an indication that the majority of the wind energy is used to lift the plasma from the Sun's gravitational well. 
Figure~\ref{fig:f5} shows scatter plot of the mass loss rate as a function of the solar wind kinetic energy rate, both on a logarithmic scale, 
for data taken near 1~AU and by Ulysses during 1996 and 2000, as well as similar plot for long-term solar wind data taken between 1965-2010 
(monthly averages), and Voyager II data taken between 1980-2009. The slope for all plots ranges between 0.5-1, suggesting that indeed, a more 
massive parcel of solar wind plasma has more energy. The spread and dynamic range of the plots is clearly related to the 
steadiness of the wind. The fast wind, observed by Ulysses during solar minimum covers a very narrow range, while the range is wider for slow 
wind observed near 1~AU during solar minimum. The range gets even wider for observation of the slow solar wind taken both by Ulysses and near 
1~AU during solar maximum. Naturally, the long-term data taken at 1~AU shows similar trend. The widest range appears in the Voyager II data, 
where in fact it is composed of two populations. One is the solar wind and the other is the gas measured at the Heliosheath.


\section{Discussion}
\label{sec:Discussion}

Based on the results presented in Section~\ref{sec:Res}, the mass flux carried by the solar wind does not have an obvious dependency 
on stellar activity. As mentioned in Section~\ref{sec:Intro}, this can be explained by assuming that the solar mass loss rate to the solar wind is 
determined by the amount of open magnetic flux opened up to the heliosphere. In contrast, solar activity level is determined by the magnetic flux maintained in the closed 
loops. The order of magnitude variations of the closed flux through the solar cycle are much greater than the factor of 2 variations of solar open flux 
\citep{Wang00}. The small variations in the open flux are due to the fact that during solar minimum, most of the open flux originates from the large polar coronal holes, 
while during solar maximum, much of it originates from the vicinity of active regions \citep{Schrijver03}. The amount of open flux is also modified by the level of magnetic 
reconnection between closed and open flux \citep{Fisk05}. This effect is enhanced during solar maximum due to the complicated field structure, where the reconnection is driven by 
differential rotation, heliospheric current sheet tilt angle from the solar rotation axis, and CMEs which carry closed flux and force it to reconnect with the heliospheric 
open flux  \citep{Fisk96,Crooker02,Owens08,Owens11,Yeates10}. As with the solar open flux, the mass loss rate of the solar wind is not constant, but is scattered around an average value 
of about $2-4\cdot10^{-14}\;M_\odot\;yr^{-1}$. Variations in the mass flux do not depend on variations in solar activity, but they are mostly due to density variations at the 
wind source region.  

Suppose the mass loss rate of a sun-like star is determined by the amount of open stellar magnetic flux. Then we can assume that the wind 
mass flux is determined at the Alfv\'en surface, where the stellar wind speed is equal to the Alfv\'en speed, and all magnetic field lines are open. In addition, the mass 
flux carried by the stellar wind needs to escape the gravitational well of the star, i.e. reach the stellar escape velocity, $u_e=\sqrt{2GM_\star/R_\star}$, where $G$ is the 
gravitational constant, and $M_\star$ and $R_\star$ are the stellar mass and radius, respectively. The solar escape velocity is $u_{e\odot}=617\;km\;s^{-1}$. 
Based on these assumptions, Table~\ref{table:t1} shows electron number densities and solar mass loss rates, $\dot{M}_\odot=n_em_pu_{e\odot}\cdot 4\pi r^2$, as a function 
of height above the solar surface near helmet streamers and in coronal holes, taken from \cite{Guhathakurta96}. It can be seen that all values for the solar mass loss rates are 
within the range of  the observations. The values are higher than the value of  $2\cdot10^{-14}\;M_\odot\;yr^{-1}$ since the actual solar Alfv\'en surface is probably higher than 
$5R_\odot$, where the density is lower.

To derive a more general expression for stellar mass loss rates, we can simply equalize the squares of 
the Alfv\'en and escape speeds at the Alfv\'en surface to obtain:
\begin{equation}
u^2_e=u^2_A=\frac{B^2(r_A)}{4\pi\rho(r_A)},
\end{equation}
or alternatively
\begin{equation}
B^2(r_A)=4\pi\rho(r_A)u^2_e=\frac{\dot{M}u_e}{r^2_A},
\end{equation}
with $\dot{M}=4\pi r^2_A\rho(r_A)u_e$ being the mass loss rate through the Alfv\'en surface, assuming it is a sphere. 
Defining $d$ as the average distance of the Alfv\'en surface in stellar radii ($r_A=dR_\star$), and assuming that the magnitude of $B$ falls like $d^{-3}$ so that $B=B_0d^{-3}$, 
we can finally obtain an expression for the stellar mass loss rate as a function of the magnetic and gravitational stellar parameters:
\begin{equation}
\dot{M}=\frac{B^2_0R^2_\star}{d^4u_e}.
\label{Mdot}
\end{equation}
The definition and interpretation of  $B_0$ can be one of the following. 1) The weak, continuos magnetic field on the stellar surface as observed on the Sun by \cite{Schrijver11}; 
2) the weak dipole component of the stellar field; or 3) the "floor" value of the open magnetic flux \citep{Owens08}. For the solar case, any choice will set $B_0$ to be of the order of 
$2-5\;G$. With the choice of $d$ ranging between $2-10R_\odot$, and solar escape velocity, we obtain a solar mass loss rate 
ranging between $10^{-15}-10^{-13}\;M_\odot\;yr^{-1}$ - a range that is in agreement with observations.

\begin{figure*}
\centering
\includegraphics[width=3.in]{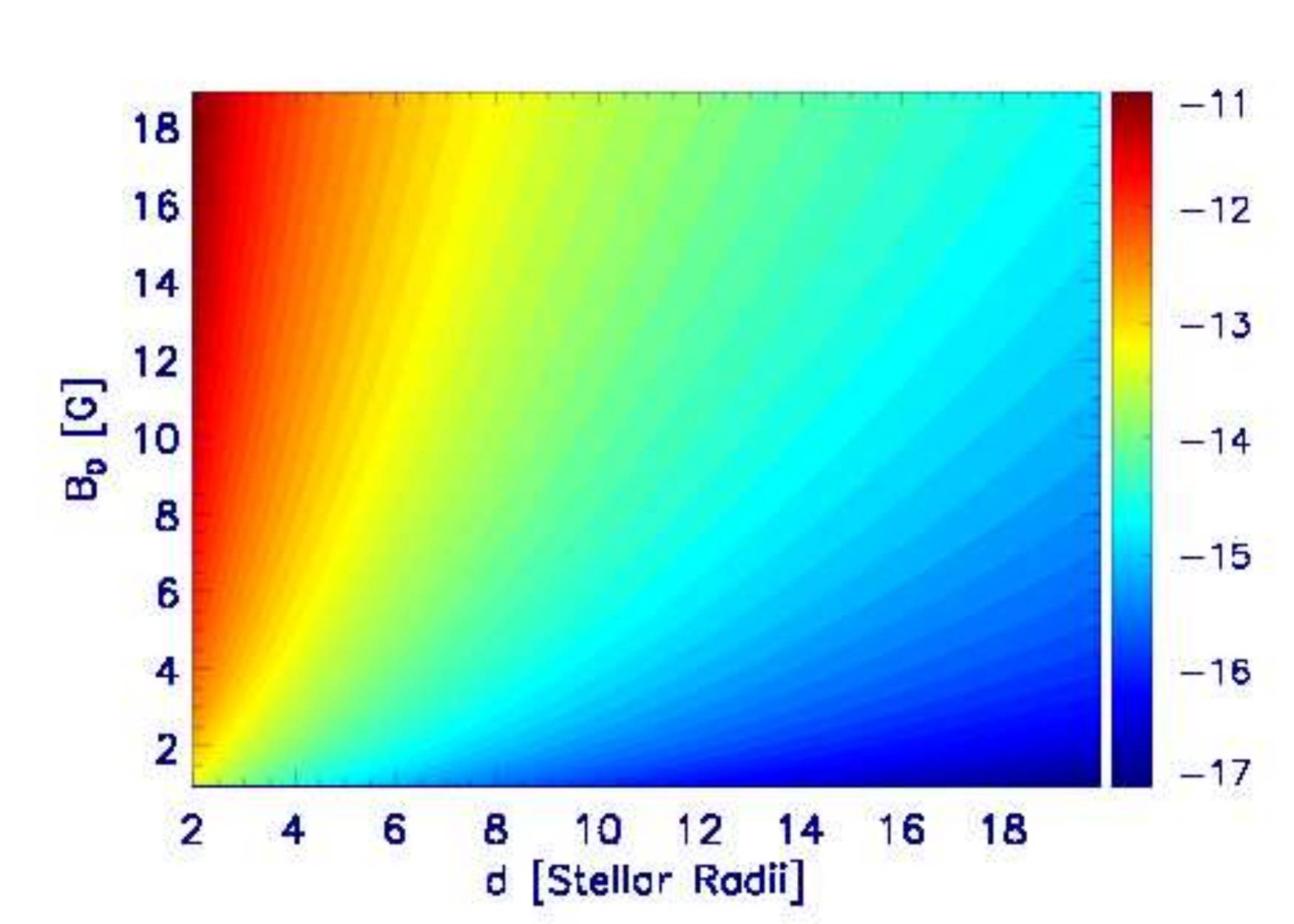}
\includegraphics[width=3.in]{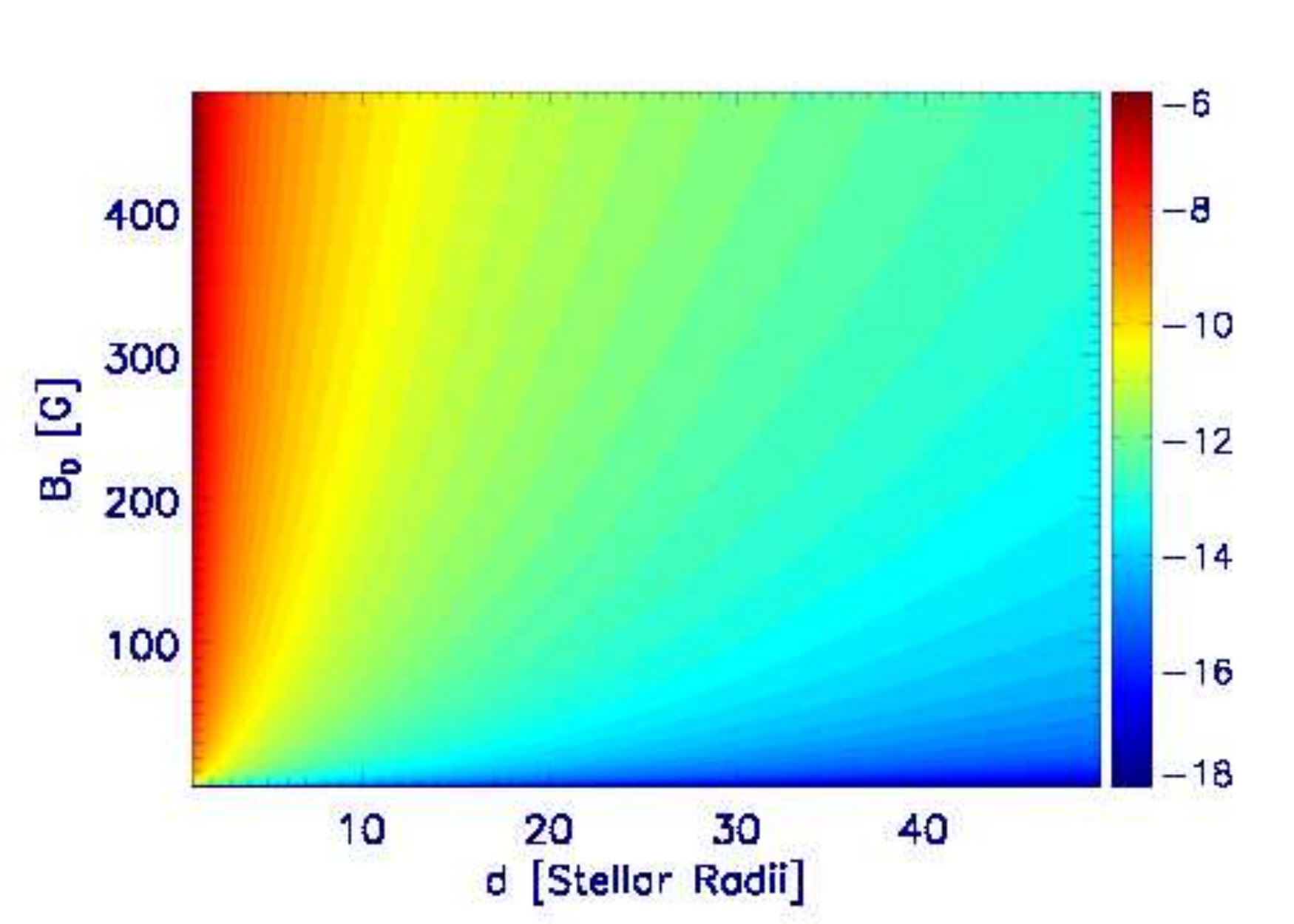} \\
\includegraphics[width=3.in]{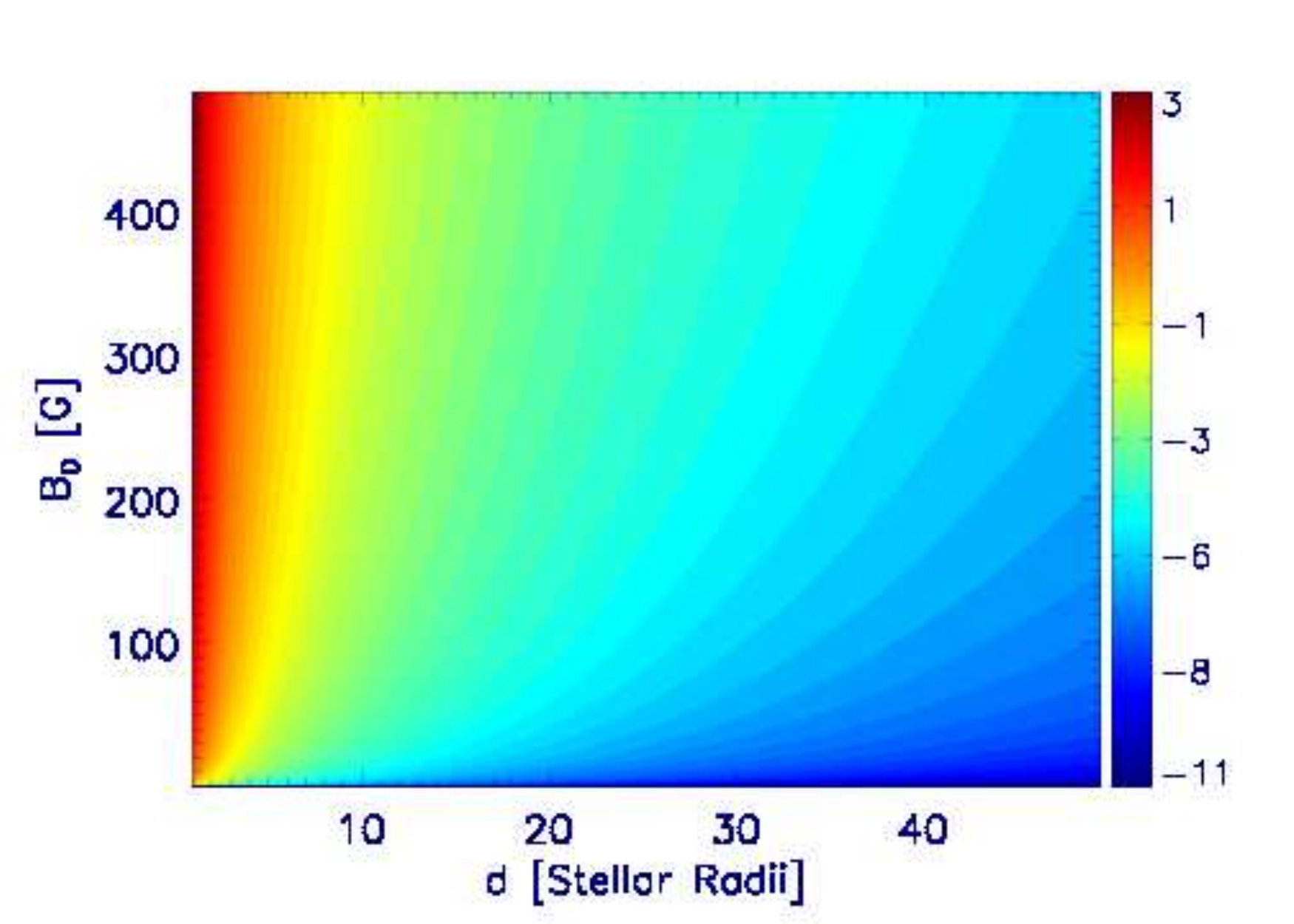}
\caption{Top: logarithm of the stellar mass loss rate based on Eq.~\ref{Mdot} for smaller values of $B_0$ and $d$ (left) and for larger values of $B_0$ and $d$ (right). 
Bottom: logarithm of $p_B$ in $dyne\;cm^{-2}$ as a function of $B_0$ and $d$.}
\label{fig:f6}
\end{figure*}

For Sun-like stars, $B_0$ and $d$ are unknown. The value of $M_\star/R_\star$ is about the same for most stars and does not exceed 2.5 for B and O type stars 
\citep{BinneyMerrifield98}. Therefore, stellar escape velocities of Sun-like stars are limited to a value that is of the order of $600\;km\;s^{-1}$. The two upper panels in Figure~\ref{fig:f6} show 
the stellar mass loss rate as a function of $B_0$ and $d$ (based on Eq.~\ref{Mdot}), using solar escape velocity. The upper-left panel shows the mass loss rate for values of $B_0$, which are up to about order of magnitude larger 
than the solar value. It can be seen that most mass loss rates are in the range $10^{-15}-10^{-13}\;M_\odot\;yr^{-1}$. The rest of the plot covers unrealistic values, since $d$ 
gets larger as $B_0$ increases. The upper-right panel shows similar plot for much stronger $B_0$ and larger value of $d$, which represent young active stars. In this plot, the 
mass loss rates range between $10^{-18}-10^{-6}\;M_\odot\;yr^{-1}$. 

To consider a realistic range, one should keep 
in mind that the extreme strong fields of up to few Kilo-Gausses observed on active stars are rather local, and are not the average weak component nor the dipole field. In order for the stellar wind 
to open the magnetic field lines and allow the mass to escape, the dynamic pressure of the wind, $p_d=\rho u^2$, must overcome the field's magnetic pressure, $p_B=B^2_0/8\pi d^6$. 
In the solar case, the dynamic pressure of the solar wind near the Alfv\'en surface is of the order of $10^{-5}-10^{-4}\;dyne\;cm^{-2}$. The lower plot of Figure~\ref{fig:f6} shows the magnetic pressure as a 
function of $B_0$ and $d$ for the strong field range. It can be seen that the typical solar dynamic pressure of $10^{-4}\;dyne\;cm^{-2}$ is not seen below $d=10R_\star$. Therefore, even for stronger fields, realistic mass loss rates are within the range of $10^{-15}-10^{-12}\;M_\odot\;yr^{-1}$.

\section{Conclusions}
\label{sec:Conclusions}

I study the dependency of solar mass loss rate on the level of solar activity and solar X-ray flux in order to test whether X-ray flux can be a good proxy for stellar mass loss rates. 
I exclude short-term transient and spikes from the solar wind and solar X-ray data, in order to better estimate the ambient values of the solar mass loss rate and solar X-ray flux at 
a particular period of time. I use solar wind data taken at different position in the solar system during different periods of solar activity. 

Observations show that the solar mass loss rate 
is scattered around the value of $2\cdot10^{-14}\;M_\odot\;yr^{-1}$ with variations of a factor of 2-5, while the solar X-ray flux is modified by an order of magnitude or even higher 
through the solar cycle. The scatter of the solar mass loss rate is wider for the slow wind and is more narrow for the fast wind, since the slow wind is more sporadic than the rather 
steady fast wind. The relative constancy of the solar mass loss rate is due to the fact that it is determined by the rather constant solar open magnetic flux. Solar activity is governed 
by the closed flux, which changes much more dramatically through the solar cycle. 

I derive a simple expression for stellar mass loss rates as a function of the stellar ambient weak 
magnetic field, the stellar radius, the stellar escape velocity, and the average height of the Alfv\'en surface. This expression predicts stellar mass loss rates of 
$10^{-15}-10^{-12}\;M_\odot\;yr^{-1}$ for Sun-like stars.

\section*{Acknowledgments}

I would like to thank an unknown referee for his/her review. I thank Jeremy Drake, Vinay Kashyap, Steve Saar, Steven Cranmer, and Justin Kasper for their useful comments and discussion during the preparation of this article. 
OC is supported by SHINE through NSF ATM-0823592 grant, and by NASA-LWSTRT Grant NNG05GM44G.

\bsp




\begin{table}
\caption{Mass Flux Trough the Source Surface}
\begin{tabular}{ccc}
\hline
Distance [$R_\odot$] & $n_e\;[cm^{-3}]$ & $\dot{M}_\odot\;[M_\odot\;yr^{-1}]$\\
\hline
3	& 1.00E+06 & 8.56E-13 \\
4	& 3.00E+05 & 4.56E-13 \\
5	& 1.00E+05 & 2.38E-13 \\
\hline
Distance [$R_\odot$] & $n_e\;[cm^{-3}]$ & $\dot{M}_\odot\;[M_\odot\;yr^{-1}]$\\
\hline
3	& 1.00E+05	& 8.56E-14 \\
4	& 3.00E+04	& 4.56E-14 \\
5	& 2.00E+04	& 4.75E-14 \\
\hline
\end{tabular}
\label{table:t1}
\end{table}

\label{lastpage}

\end{document}